\documentclass[twoside]{article}
\usepackage{fleqn,espcrc2}

\usepackage{graphicx}
\usepackage[figuresright]{rotating}

\newcommand{\AmS}{{\protect\the\textfont2
  A\kern-.1667em\lower.5ex\hbox{M}\kern-.125emS}}

\title{Fixed Number and Quantum Size Effects in Nanoscale Superconductors}

\author{K. Tanaka and F. Marsiglio,
\address{Department of Physics, University of Alberta,
Edmonton, Alberta, Canada T6G 2J1}}
       
\begin{document}

\begin{abstract}
In recent experiments on nanoscale Al particles,
whose electron number was fixed by charging effects,
a ``negative gap'' was observed in particles with an odd number of
electrons.
This observation has called into question the use of a grand canonical
ensemble
in describing superconductivity in such ultrasmall particles.

We have studied the effects of fixed electron
number and finite size in nanoscale superconductors, by applying  
the canonical BCS theory for the attractive Hubbard model.  
The ground state energy and the energy gap 
are compared with the conventional and parity-projected grand canonical 
BCS results, and in one dimension also with the exact solutions 
by the Bethe ansatz.
The crossover from the bulk to quantum limit is studied 
for various regimes of electron density and coupling strength.
The effects of boundary conditions and different lattice structures
are also examined.

A ``negative gap'' for odd electron number emerges 
most naturally in the canonical scheme.  
For even electron number,
the gap is particularly large for ``magic numbers'' of electrons 
for a given system size or of atoms for a fixed electron density.
These features are in accordance with the exact solutions,
but are essentially missed in the grand canonical results.
\vspace{1pc}
\end{abstract}

\maketitle

The ability to fabricate ultrasmall superconducting particles in a
reasonably controlled way \cite{ralph95} allows us to revive 
old questions \cite{anderson59}. The question we focus on here
is the validity (and usefulness) of the grand canonical ensemble
vs a canonical one for a description of very small superconducting
particles.

The canonical and parity-projected BCS formalisms have been described 
elsewhere \cite{tanaka99,braun98,janko94}.
Figure 1 shows a comparison of even and odd canonical (CBCS) and
grand canonical (GCBCS) ground state energies, 
along with exact ones, for the attractive Hubbard model (AHM) in 1 D,
for coupling strength 
(scaled by the kinetic energy parameter $t$) $|U|/t=4$ and 10.
Odd-even effects are clearly discernible.

Figure 2 shows the (even) GCBCS result for the gap ($\Delta_{\circ}$)
vs. electron density $n$, along with the smoothed density of states (DOS) 
as a function of single-electron energy $\epsilon_k$.
The structure visible in the gap
requires painstaking numerical work, and reflects the underlying
discrete density of states (as seen in Fig. 3 below). In Fig. 3 we
show a smaller system, with both CBCS and (even) GCBCS results as a function 
of electron density. Quite a few anomalously high gaps occur, at various
values of $n$, as revealed by the CBCS result ($\Delta_{N_e}$).

Finally, in Fig. 4 we examine the CBCS gap (normalized by the bulk value)
as a function of system size $N$, 
for systems with an even number of electrons. 
There are two distinct curves
which approach the bulk limit (solid circles), corresponding to
$N_e = 4m$ or $N_e = 4m+2$, with $m$ an integer, the so-called `super-even'
effect \cite{tanaka99}. Clearly, the transition to the bulk is smooth.
\begin{figure}[htb]
\begin{center}
\includegraphics[width=12pc,height=15pc]{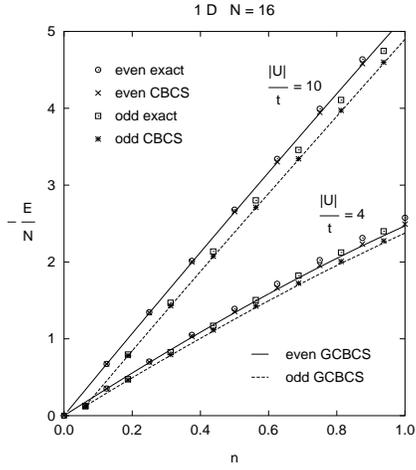}
\caption{Ground state energy vs. $n$ for even and odd electron numbers.}
\label{fig1}
\end{center}
\end{figure}
\begin{figure}[htb]
\begin{center}
\includegraphics[width=14pc,height=18pc]{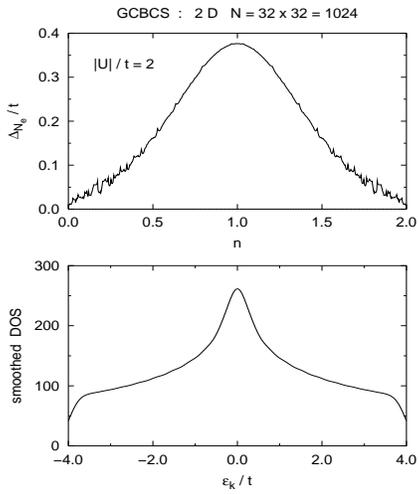}
\caption{GCBCS (even) gap function vs. n (upper graph) and the
single-electron DOS vs. $\epsilon_k$  (lower graph) in 2D. Structure in 
upper graph is not noise, i.e., it is due to discrete level structure.}
\label{fig2}
\end{center}
\end{figure}
\begin{figure}[htb]
\begin{center}
\includegraphics[width=14pc,height=18pc]{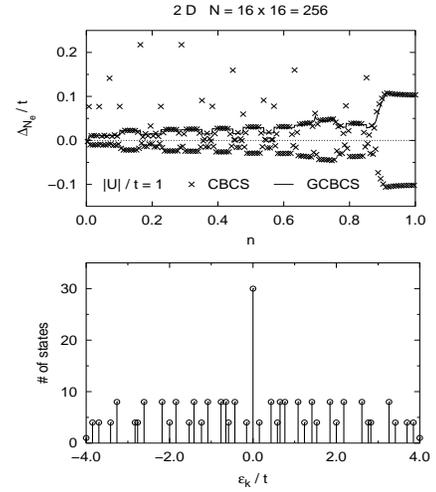}
\caption{CBCS vs. GCBCS (even) gap vs. $n$ (upper graph) and discrete
DOS (lower graph).}
\label{fig3}
\end{center}
\end{figure}
\begin{figure}[htb]
\begin{center}
\includegraphics[width=14pc,height=17pc]{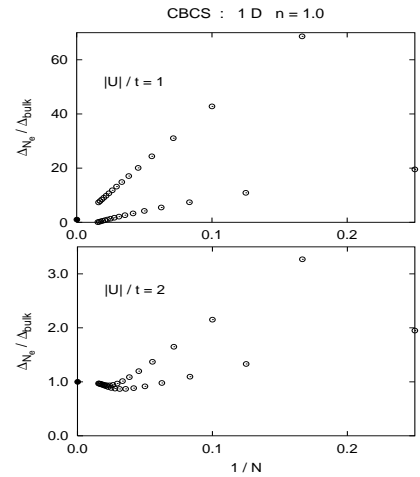}
\caption{Canonical gap (for even electron numbers only) vs. $1/N$.
Bulk-limit results are shown by solid circles at $1/N=0$.}
\label{fig4}
\end{center}
\end{figure}

\end{document}